\RequirePackage{fix-cm}
\documentclass[smallextended]{svjour3}       
\smartqed  
\usepackage{graphicx}
\usepackage{gensymb} 	

\usepackage{natbib}
\input aasmacros.sty 	
\def\asr{\reff@jnl{Advances in Space Research}}

\def\cxb{cosmic X-ray background}

\def\wfi{wide-field hard X-ray imager}
\def\Wfis{Wide-field hard X-ray imagers}
\def\wfis{wide-field hard X-ray imagers}

\begin{document}

\title{Designing large pixelated CdTe detection planes for hard X-ray transients detection
}

\titlerunning{Designing large pixelated CdTe detection planes}        

\author{Karine Lacombe        \and       
Carine Amoros \and 
Jean-Luc Atteia \and
Armelle Bajat \and 
Laurent Bouchet \and
Jean-Pascal Dezalay \and 
Philippe Guillemot \and
Baptiste Houret \and 
Fran\c cois Lebrun \and
Sujay Mate \and
Roger Pons \and
Henri Triou \and
Vincent Waegebaert \and
}


\institute{K. Lacombe,  C. Amoros, J-L. Atteia, L. Bouchet, J.-P. Dezalay, B. Houret S. Mate, R. Pons, V. Waegebaert: IRAP, Universit\'e de Toulouse, CNES, CNRS, UPS, (Toulouse), France \\
A. Bajat: CEICO, Institute of Physics, Czech Academy of Sciences, Prague, Czech Republic \\
P. Guillemot: CNES, 18 Avenue Edouard Belin, Toulouse Cedex 9,31401, France \\
F. Lebrun: APC, AstroParticule et Cosmologie, Universit\'e Paris Diderot, CNRS/IN2P3, CEA/Irfu, Observatoire de Paris Sorbonne, Paris, France \\
H. Triou: CEA Saclay Ð DSM/Irfu/D\'epartement dÕAstrophysique, France
}

\date{Received: date / Accepted: date}

\maketitle

\begin{abstract}\
We discuss the need for very large detection planes for the detection of hard X-ray transients in the multi-messenger era that started with the quasi-simultaneous detection of GRB~170817A by \textit{Fermi}/GBM and \textit{INTEGRAL}/SPI on one hand and the gravitational waves event GW 170817, detected by the LVC collaboration, on the other hand. 
Then, we present a brief survey of current and future instruments and their expected sensitivity, pointing the fact that the gain in the number of GRBs is achieved by future projects thanks to their larger field of view rather than to their larger effective area. 
Based on our experience with \textit{SVOM}/ECLAIRs, we then address various problems associated with the realization of very large detection planes (\mbox{$\ge 1 \mathrm m^{2}$}), and we demonstrate that CdTe detectors are well suited for this task. 
We conclude with a discussion of some key parameters that must be taken into account in the realization of instruments based on these detectors.
We hope that this paper will motivate the elaboration of detailed proposals of large area wide-field hard X-ray monitors that will be crucially needed in the next decade.

\end{abstract}


\section{Introduction}
\label{sec_intro}
The study of cataclysmic cosmic phenomena is a key endeavor of $21^{st}$ century astronomy, fostered by the advent of new instruments capable of detecting cosmic explosions routinely over large regions of the sky.
While such studies have long been the prerogative of X-ray and $\gamma$-ray astrophysics, the advent of new astrophysical messengers such as gravitational waves and neutrinos open new windows on energetic cosmic phenomena: the collapse of massive stars, the mergers of compact objects or violent accretion episodes on galactic and extragalactic black holes.
In the electromagnetic domain, optical and radio observatories can now monitor a significant fraction of the sky, allowing the efficient detection of the electromagnetic counterparts of these violent events.

\begin{figure}[h]				
\centering
\includegraphics[width = 0.85\textwidth]{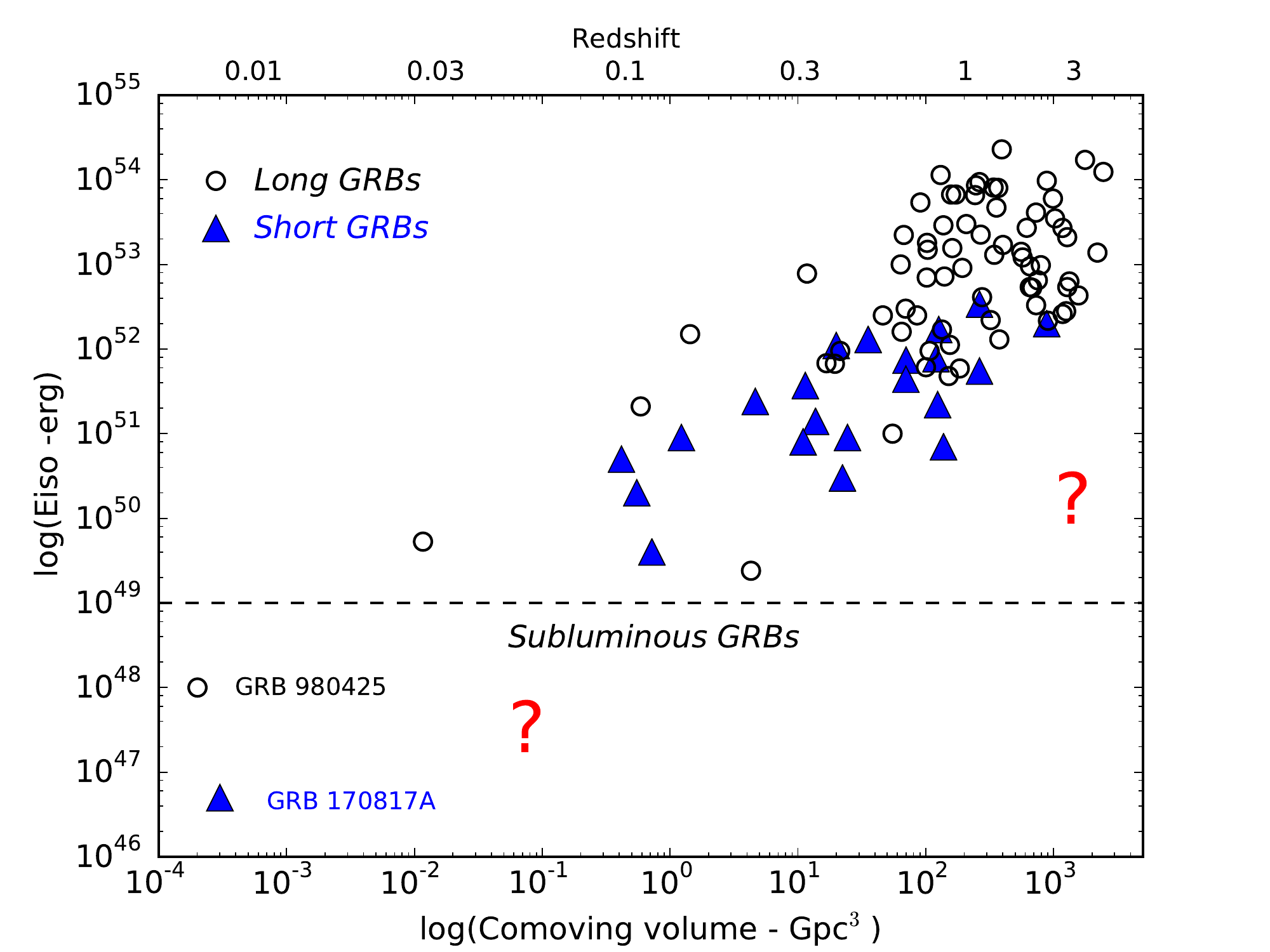}
\caption{Distribution of selected GRBs in the plane log(Volume) - log($\mathrm{E_{iso}}$). 
Long GRBs are taken from \cite{Amati2008} and short GRBs from \cite{Fong2015}. 
The values for GRB~980425 are from \cite{Galama1998} and for GRB~170817A from \cite{Abbott2017} and \cite{Goldstein2017}. 
Two regions where the GRBs are too faint to be detectable with the present instrumentation are indicated with red question marks.
}
\label{fig_grbvz} 
\end{figure}

This new panorama considerably widens the range of rare phenomena that can be studied in detail.
While the transient hard X-ray sky exhibits very energetic phenomena such as GRBs, which are detectable up to high redshifts, it also encompasses less energetic events, only detectable up to few hundred Mpc with the current instrumentation. 
This is the case of subluminous GRBs (see Fig. \ref{fig_grbvz}), of SNe shock breakout and events similar to GRB~170817A \citep{Goldstein2017} emitted by two merging neutron stars that produced the gravitational wave transient detected by the LVC collaboration as GW~170817 \citep{Abbott2017}. 
The detection of such events in large numbers in hard X-rays calls for a significant increase of sensitivity in the hard X-ray range. 
This is illustrated in Fig.~\ref{fig_grbvz}, that shows the extent of uncharted territories for both subluminous GRBs and classical (long and short) GRBs. 
The situation of subluminous GRBs is particularly crucial in view of the planned increase of sensitivity of gravitational wave interferometers, which will have by 2022 the sensitivity to detect binary neutron star mergers in a volume larger than \mbox{$10^{-2}$ Gpc$^3$} \citep{Abbott2018}.
The study of the counterparts of GW events, however, encompasses two different needs. 
First, to measure accurately the prompt and delayed electromagnetic spectrum over a band as large as possible for every GW event to cast some light on the physics at work in the source.
This requires a sensitive wide band spectrometer watching the entire sky (e.g. INTEGRAL-SPI ACS).
Second, to provide as soon as possible an arcminute localisation of the GRBs to allow for follow-up observations.
This requires a wide FOV coded mask telescope equipped with a sensitive imager, this paper is devoted to this need.

While new technologies are emerging for the construction of highly sensitive X-ray monitors with large field of views, such as the lobster eye optics proposed for Einstein Probe \citep{Yuan2017} or THESEUS \citep{Amati2018}, the most successful current technology is based on large CdTe (Cadmium Telluride) or CZT (Cadmium Zinc Telluride) detection planes looking at the sky through coded masks.
This method has been used by several missions from the early times of hard X-ray astronomy (see for instance \cite{Cavallari2017} for a review), and is currently used by ISGRI onboard \textit{INTEGRAL} \citep{Lebrun2003}, by \textit{Swift}/BAT \citep{Barthelmy2005} or by \textit{AstroSAT}/CZTI \citep{Vadawale2016}. 
Tables \ref{tab_current} and \ref{tab_future} provide a list of current and future hard X-ray imagers using coded-mask technology (not restricted to CdTe detectors in Table 2).
These tables show that the detection planes of current and future instruments have areas of hundreds to thousands square centimeters, with a maximum of 5200~cm$^2$ for the Burst Alert Telescope (BAT) onboard the Neil Gehrels \textit{Swift} Observatory (hereafter called \textit{Swift}).

In this paper we use the lessons that we have learned during the development of the detection plane of ECLAIRs for the \textit{SVOM} mission to discuss the requirements and constraints on large CdTe detection planes for future hard X-ray imagers.
Even if the final performance of an instrument depends on many of its characteristics such as e.g. its geometry (including the coded mask and the shield) or its orbit, we are focusing here on the detection plane because it lays the foundations of the instrument performance.




This paper is organized as follows. 
In section \ref{sec_detection} we discuss the sensitivity requirements for future wide-field hard X-ray monitors, and a figure of merit for the detection of short transients with this type of instruments.
Section \ref{sec_status} presents current or planned instruments designed to survey the X-ray sky, which are compared according to the figure of merit previously defined.
In section \ref{sec_eclairs}, we discuss the lessons learned during the construction of the detection plane for \textit{SVOM}/ECLAIRs, and their consequences for the design of large CdTe detection planes.
Finally, in section \ref{sec_discussion} we propose simple guidelines for the design of future instruments.
Throughout this paper, we consider an energy domain ranging from few keV to few hundred keV, which we call the hard X-ray range.

\section{The detection and localization of cosmic transients}
\label{sec_detection}

Most high-energy transients are unpredictable, they occur at any time in any direction of the sky and their durations and spectra are highly variable.
Thus, high-energy monitors must have a wide field of view (FoV), cover a broad energy range and survey the sky for transients of all durations (from few milliseconds to hours).
In this section we discuss how the instrumental characteristics impact the number of detected transients.

\subsection{The background}
\label{sub_background}

\begin{figure}			
\centering
\begin{minipage}{.48\textwidth}
\centering
\includegraphics[width = 0.95\textwidth]{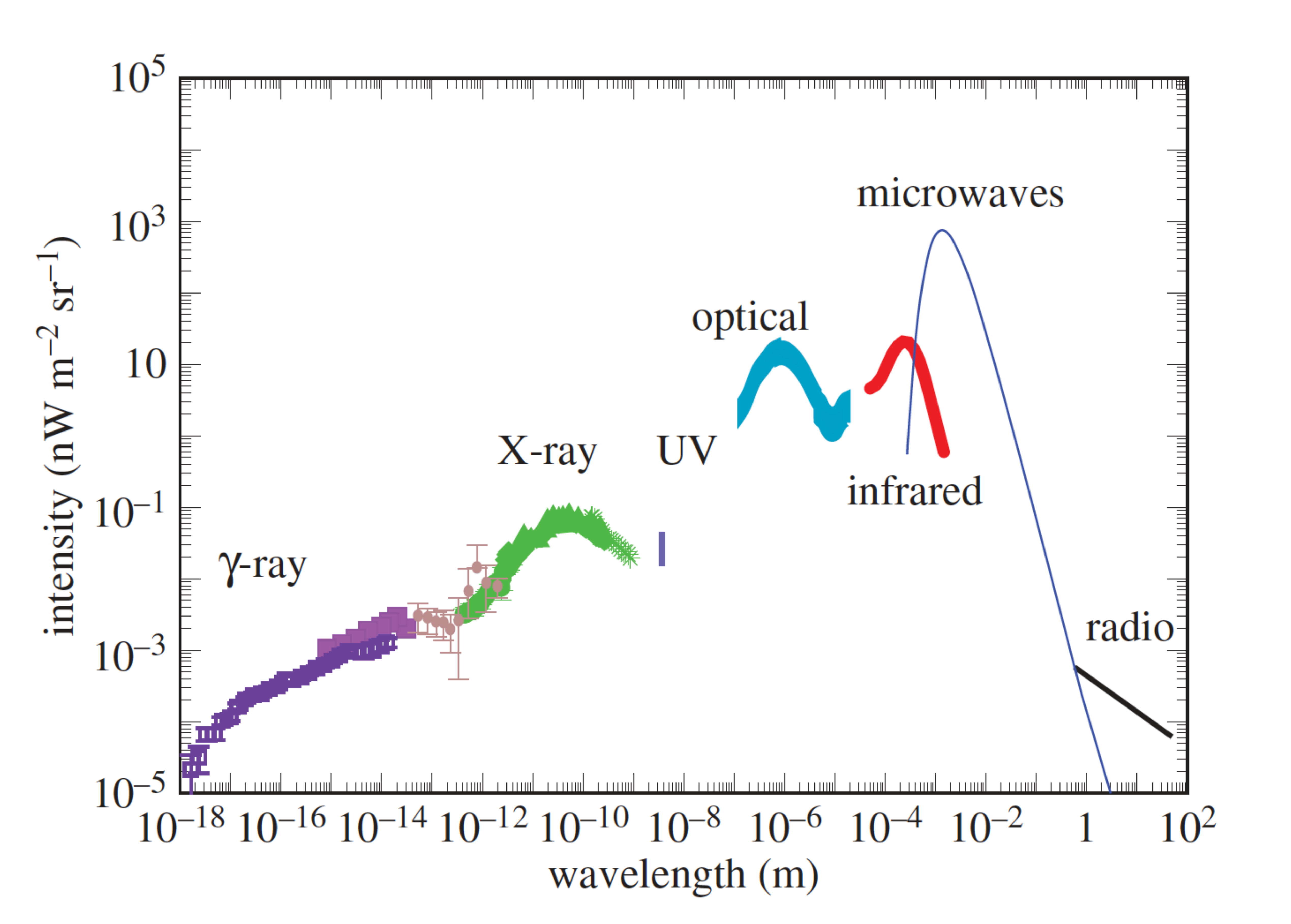}
\end{minipage}
\begin{minipage}{.48\textwidth}
\centering
\includegraphics[width = 0.95\textwidth]{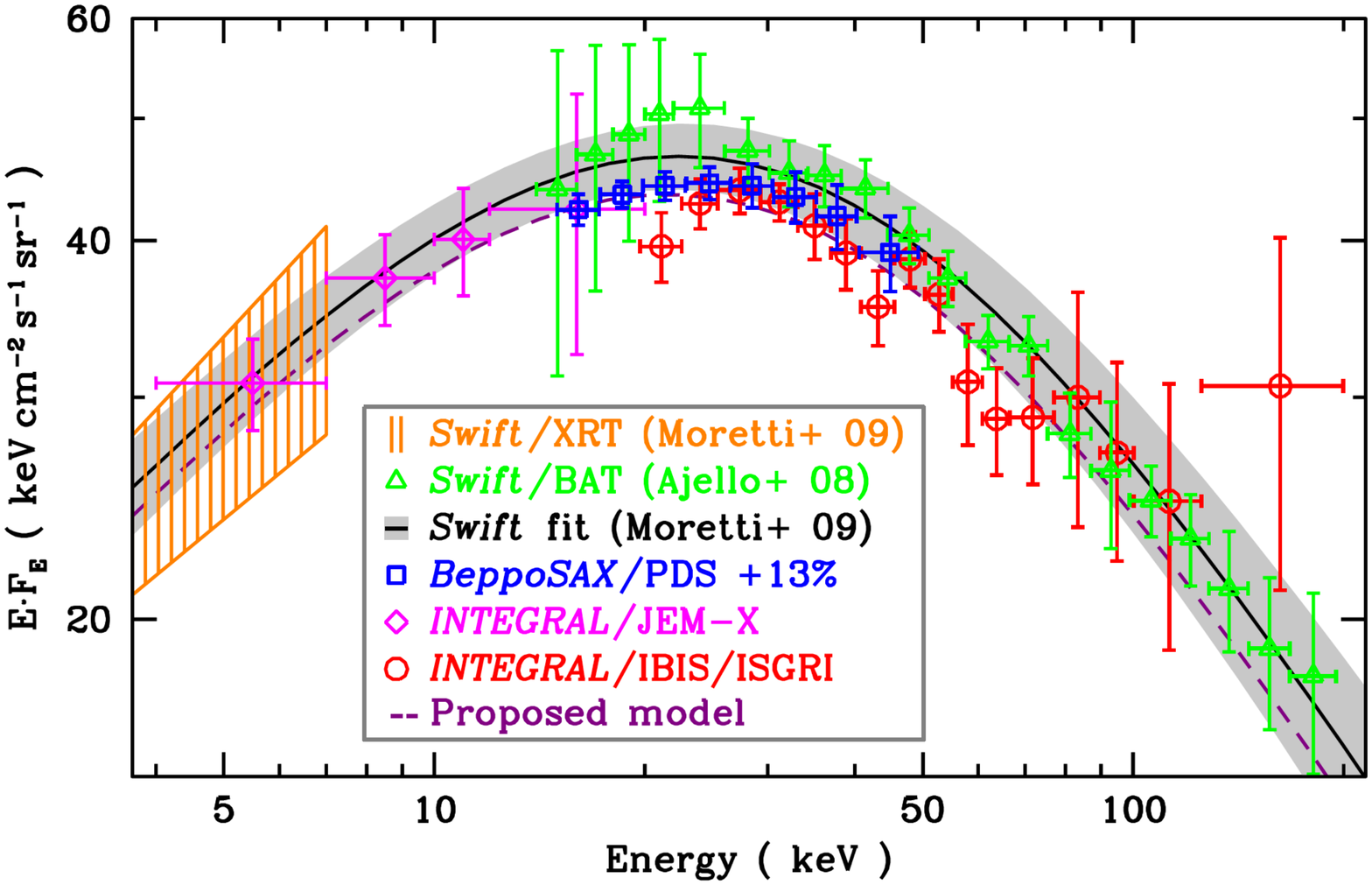}
\end{minipage}
\caption{\textit{Left panel}. Intensity of the Extragalactic Background Light (EBL) across the electromagnetic spectrum (from \cite{Cooray2016}).
The \cxb\ is shown in green.
\textit{Right panel}. Detailed measurements of the cosmic X-ray background, which is the main source of background for wide-field instruments operating below 100 keV (from \cite{Turler2010}).}
\label{fig_cxb}
\end{figure}

The sensitivity of \Wfis\ is limited by the large background of these instruments which collect numerous photons from the cosmic X-ray background (CXB, Fig. \ref{fig_cxb}), mostly due to extragalactic AGNs.
The CXB is the dominant source of background for wide field instruments observing in the range [10--100]~keV.
According to \citet{Turler2010}, the photon flux from the CXB typically amounts to \mbox{6.9 ph cm$^{-2}$ s$^{-1}$ sr$^{-1}$} (resp. \mbox{3.7 ph cm$^{-2}$ s$^{-1}$ sr$^{-1}$}) in the energy range [5 -- 150]~keV (resp. [10 -- 150]~keV), representing few thousand background counts per second for instruments with an effective area of several hundred square centimetres and a field of view of few steradians.

\begin{figure}[]				
\centering
\includegraphics[width = 0.85\textwidth]{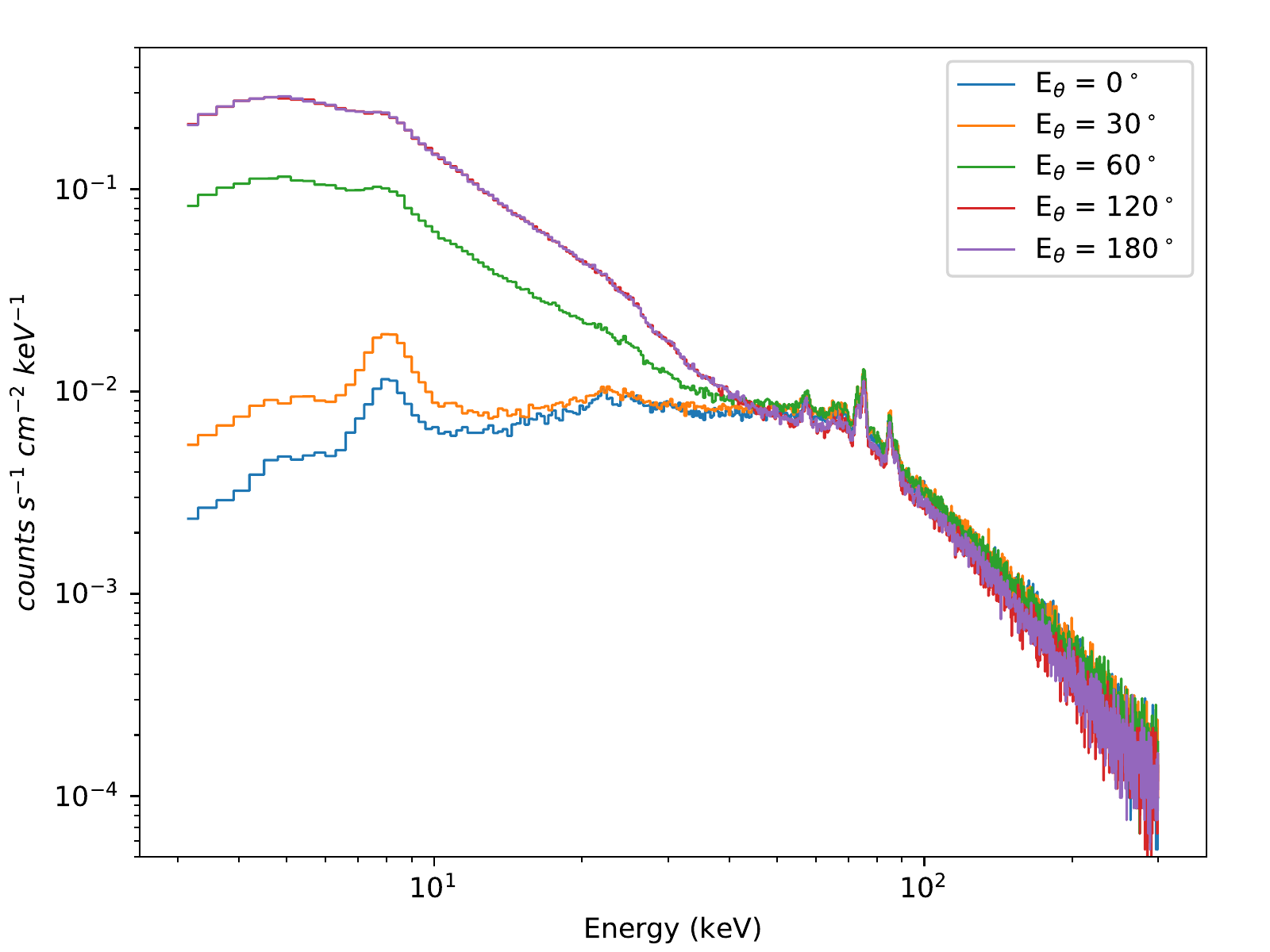}
\caption{Simulated background spectrum in the energy range 4-150 keV, for various positions of the Earth with respect to the pointing direction of ECLAIRs. The cosmic X-ray background, which is occulted by the Earth, is the dominant source of background below \mbox{$50-60$ keV}.
Fluorescence lines from copper (\mbox{E $\approx 8$~keV}), tantalum and lead (\mbox{E $\approx 70$~keV}) in the mask and shield appear clearly when the Earth blocks the CXB, they will be used for in-flight calibration.  
}
\label{fig:eclairscxb}
\end{figure}

Another important contribution to the background is due to cosmic rays and to charged particles from the Earth radiations belts which interact with the materials of the satellite and produce prompt and delayed secondary photons and charged particles which deposit all or part of their energy in the detectors.
In the following, we call particle background this contribution.
At low energies, the CXB photons dominate the background while the particle background dominates at higher energies.
The transition between these two background regimes depends on the geometry of the instrument, for \textit{SVOM}/ECLAIRs it is located around 60~keV.
The importance of the CXB is illustrated in Fig.~\ref{fig:eclairscxb}, which shows simulated background spectra of ECLAIRs, when the Earth occults the field-of-view partly or completely. 

\subsection{The signal-to-noise ratio}
\label{sub_snr}
The strength of the signal received from a transient source occurring within the field of view of a hard X-ray monitor depends mostly on the effective area of the detector (\mbox{$\rm A_{\rm eff}$}) and its energy range.
The sky background, which dominates the background, B, also depends on these two characteristics, plus the field of view ($\Omega$): \mbox{$B \propto A_{\rm eff}  \times \Omega  $}.
Overall, for a given energy range and time window, the signal-to-noise ratio (SNR) of a transient source is determined by the field of view and the effective area of the instrument, and we can write:

\begin{equation}
{\rm SNR \propto {S \over \sqrt{B}} \propto {A_{\rm eff} \over \sqrt{A_{\rm eff} \times \Omega}} \propto {A_{\rm eff}^{1 \over 2} \times \Omega^{-{1 \over 2}}} }
\label{eq_1}
\end{equation}

The maximum SNR will be reached when the search is adapted to the duration of the transient and to its energy spectrum.
This condition is usually fulfilled for instruments using a trigger software that samples the data to look for significant excesses over many temporal and spectral windows.
Moreover, different values of SNR are usually required for different purposes.
For instance, the SNR which is considered sufficient to detect a transient may be insufficient to localize it.
This is especially true for short (sub-second) bursts which may be easily detectable as short temporal excesses, but difficult or impossible to localize with coded-mask instruments, due to their small number of photons. 

\subsection{The number of detected transients}
\label{sub_number}
In this section, we use the simple parametrization of Eq.~\ref{eq_1} to evaluate the number of transients detected above a given SNR threshold. 
The number of transients detected by \wfis\ depends on the instrument geometry, on its observing time, and on the ``quality'' of the field of view, which may be partly obscured by the Earth, or contaminated by galactic sources.
If we assume a clean field of view and a population of hard X-ray transients with a cumulative peak flux distribution (the logN--logP curve) following a power law of slope $\gamma$, with $\rm log(N>P)\propto P^{-\gamma}$, then the number of detected transients per unit of observing time depends on the characteristics of the instrument:

\begin{equation}
\rm N_{\rm det} \propto N(>P_{min}) \times \Omega \propto  P_{min}^{-\gamma} \times \Omega 
\label{eq_2}
\end{equation}

where P$_{\rm min}$ is the minimum peak flux detectable by the instrument.
Considering two instruments with identical SNR thresholds (e.g. 7~$\sigma$) but different sensitivities, the most sensitive instrument will attribute a higher SNR to GRBs, and detect GRBs with a lower peak flux, such as $\rm P_{min} \propto SNR^{-1}$.
Using Eq.~\ref{eq_1}, we can write:

\begin{equation}
{\rm P_{min} \propto SNR^{-1} \propto {A_{\rm eff}^{-{1 \over 2}} \times \Omega^{{1 \over 2}}} }
\label{eq_3}
\end{equation}

By replacing P$_{\rm min}$ in Eq. \ref{eq_2} We finally obtain: 

\begin{equation}
\rm N_{det} \propto A_{\rm eff}^{\gamma \over 2}  \times  \Omega^{1- {\gamma \over 2}} 
\label{eq_4}
\end{equation}

This is an order of magnitude estimate which must not be used when an accurate evaluation is needed.
Nevertheless it catches some interesting features of \wfis\ for the detection of isotropic transient sources.

First, for sources with a steep peak flux distribution (e.g. $\gamma = 1.5$), Eq. \ref{eq_4} becomes $\rm N_{det} \propto A_{\rm eff}^{0.75}  \times  \Omega^{0.25}$, showing that the size of the field of view has little or no impact on the number of detected sources.
In this situation, the gain offered by a potentially larger field of view is compensated by the loss of faint sources resulting from the increased background. 
This is true as long as the CXB dominates over the instrumental background.
For such sources, the sensitivity gains are mostly obtained by increasing the effective area.

For sources with $\gamma = 1.0$, Eq. \ref{eq_4} becomes $\rm N_{det} \propto A_{\rm eff}^{0.5}  \times  \Omega^{0.5}$, and we obtain comparable sensitivity gains by increasing the effective area or the field of view. 
The number of detected sources is directly proportional to the square root of the \textit{grasp} of the instrument, defined as \mbox{$\mathrm{G = A \times \Omega}$}, in units of \mbox{cm$^2$ sr}.

For sources with a shallow peak flux distribution ($\gamma \le 0.5$), Eq. \ref{eq_4} becomes $\rm N_{det} \propto A_{\rm eff}^{0.25}  \times  \Omega^{0.75}$, and the sensitivity gains are mostly obtained by increasing the field of view (whenever possible).
When the entire available sky is observed, the detection of new sources requires a very large increase of the effective area, since $\rm N_{det} \propto A_{\rm eff}^{0.25}$.

Second, while Eq. \ref{eq_4} provides an order of magnitude estimate of the total number of sources detectable by an instrument, it gives no information on the ``quality'' of these sources. 
For instance, an instrument designed to observe faint GRBs in high-redshift galaxies requires a large effective area but it can accommodate a small field of view (\mbox{$\sim 1$sr}).
On the other hand, an instrument looking for nearby GRBs potentially associated with gravitational waves detectable by all-sky GW interferometers on Earth, may favor a large field of view and a smaller effective area.

Finally, considering the physical limits on the field-of-view (\mbox{$ \leq 4 \pi$ sr} for the entire sky and \mbox{$ < 3 \pi$ sr} for instruments in low Earth orbit) it seems unavoidable to increase the effective area of \wfis\ in order to improve their performance in the future.
This demand is strengthened when we consider the growing interest for faint hard X-ray transients in the local universe, which are privileged sources for multi-messenger astronomy \citep[e.g.][]{Goldstein2017, Abbott2017}, but also the need to detect GRBs from the most remote regions of the universe.
Since various present days instruments use detection planes with an area of one to a few thousand square centimetres (as discussed in the next section), it is reasonable to consider the need for instruments with an effective area larger than one square meter, requiring about 2 square meters of detectors, when the mask open fraction is taken into account. 

\begin{figure}[h]				
\centering
\includegraphics[width = 0.85\textwidth]{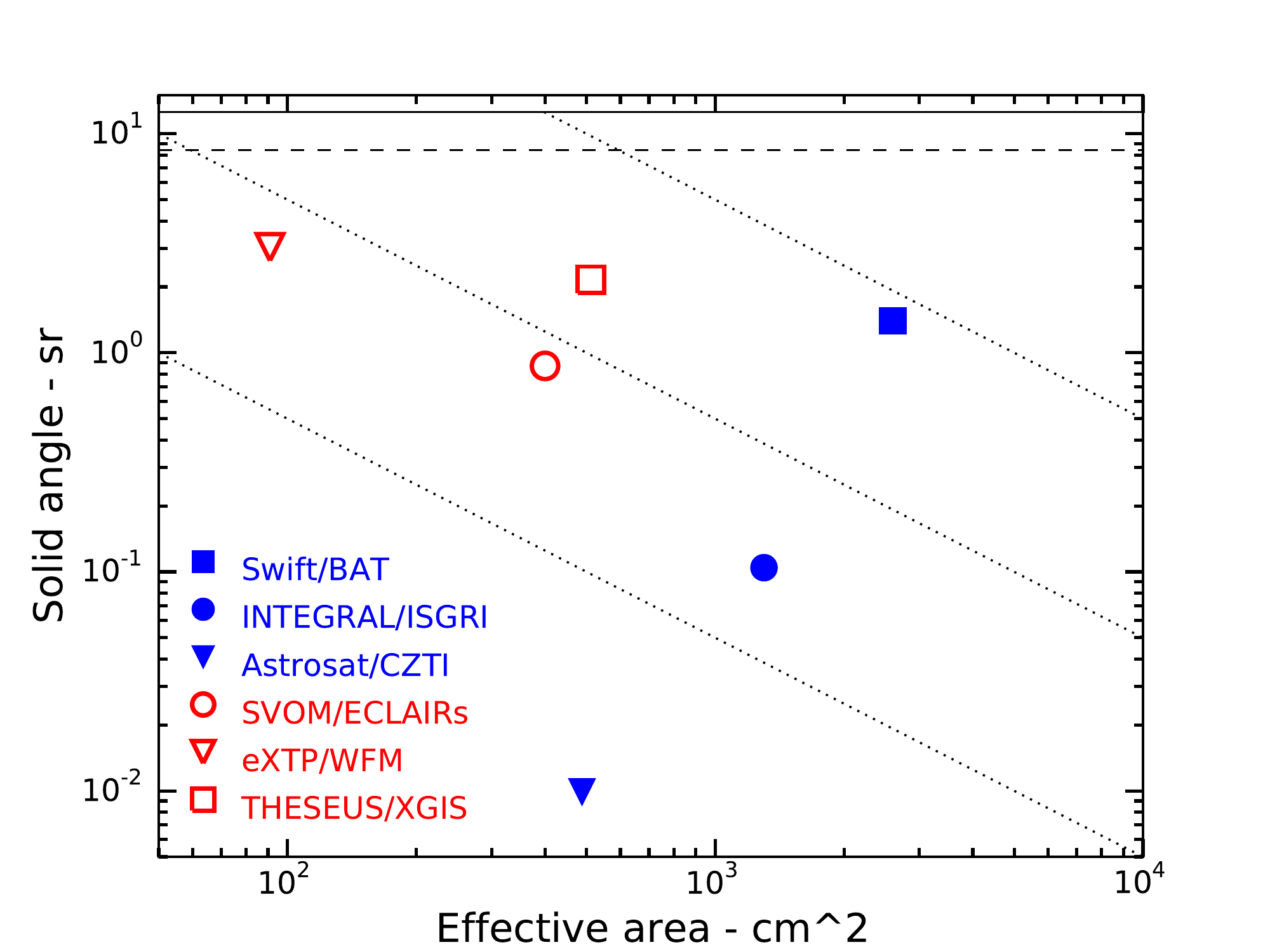}
\caption{Field-of-view (FWHM) vs effective area for various \wfis\ currently in operation or planned.
Filled blue symbols represent instruments that are in operation.
Empty red symbols represent instruments that are approved (\textit{SVOM}/ECLAIRs) or proposed ((\textit{eXTP}/WFM), \textit{THESEUS}/WFI).
The dashed lines in diagonal represent lines of equal \textit{grasp} (cm$^2$.sr), indicating instruments with the same detection power for sources with $\gamma = 1.0$, according to Eq.  \ref{eq_4}. 
The solid horizontal line is $4\pi$, the solid angle of the full sky. The dashed horizontal line is the solid angle which is not obscured by Earth for datellites in low Earth orbit at a typical altitude of 600~km.  
}
\label{fig_fom}
\end{figure}

\section{Large X-ray detection planes: a status}
\label{sec_status}

In this section, we discuss examples of hard X-ray imagers that are currently detecting and studying transient sources, based on the coded mask technology, and using CdTe or CZT detection planes such as \textit{INTEGRAL}/IBIS, \textit{Swift}/BAT or \textit{ASTROSAT}/CZTI.\\
Then we present some future instruments such as \textit{SVOM}/ECLAIRs, \textit{eXTP}/LAD and \textit{THESEUS}/XGIS-LE, comparing them with current instruments.\\

\subsection{Current instrumentation}
\label{sub_currentmissions}
\subsubsection{The ISGRI instrument onboard INTEGRAL}

The INTEGRAL ESA mission \citep{Winkler2003} is an all purposes gamma-ray observatory that has been launched in 2002 on an 72 h eccentric highly inclined orbit and is still in operation nowadays. The spectrometer SPI \citep{Vedrenne2003} features a large anticoincidence shield (ACS) that permitted the detection of the electromagnetic emission associated with GW 170817. The imager IBIS \citep{Ubertini2003} is a coded mask instrument with two detection planes: ISGRI \citep{Lebrun2003} for the low gamma-ray energies and PICsIT \citep{Labanti2003} at higher energies.
ISGRI is made of 16384 detectors read by 4096 ASICs, operating at a nominal temperature close to $0\degree$C. Its 2600 cm$^2$ area is sensitive between 15 and 1000 keV, with a maximum effective area of $1000$ cm$^2$ between 20 and 100 keV. It consists of 8 modules of 64 x 32 CdTe pixels, each one measuring \mbox{4x4 mm$^2$} and \mbox{2 mm} thick.
ISGRI provides images with 12 arcminute spatial resolution, thanks to a coded mask located 3.2 m above the detection plane. 
Its half-coded FoV is 19\degree x 19\degree, and 29.4\degree x 29.4\degree for the total field of view  \citep{Lebrun2003}. 

After 16 years of observation, with a sensitivity of $\sim3~10^{-6}$ cm$^{-2}$ s$^{-1}$ keV$^{-1}$ at 100 keV ($10^{5}$ s, 3 sigma, $\Delta E$ / E = 1/2), IBIS has detected nearly 150 bursts ($\approx9$ GRBs/yr \citep{Gotz2013}).
This relatively small number is the consequence of the rather small field of view of ISGRI (Fig. \ref{fig_fom}).

\subsubsection{The BAT instrument onboard the Neil Gehrels \textit{Swift} Observatory}
The Neil Gehrels \textit{Swift} Observatory is a satellite dedicated primarily to the detection and follow-up of GRBs, which has been launched in November 2004 into a circular low-Earth orbit with an altitude of 600 km and an inclination of 21 degrees.
The largest instrument of the Neil Gehrels \textit{Swift} Observatory is the Burst Alert Telescope (BAT), a large area \wfi\ designed to detect and localise GRBs.
\textit{Swift}/BAT provides images of the sky from 15 to 150 keV with a spatial resolution of 17 arcmin, within a field of view of 1.4 steradian. 
BAT consists of 32768 CdZnTe (CZT) semiconductor pixels of 4 by 4 by 2 mm$^3$ making a detection plane with a sensitive area of $\approx5200$ cm$^2$. 
The detectors, operating at the nominal temperature of 20\degree C, are gathered into arrays of 8 x 16 pixels, polarized to $-200$ V, and connected to a 128-channel ASIC, that is part of the readout electronics (thus needing 256 ASIC chips).

With an area of 2.7 m$^2$, the 50 \% open random mask is larger than the detection plane, which is located 1 m below it, giving a half-coded field of view of 100\degree x 60\degree (1.4 sr). 
The effective area was  \mbox{$\sim $2200 cm$^2$} at launch. 
A graded shield joins the detection plane and the mask, reducing by 95 \% the background noise due to the cosmic X-ray background. 

Since 2004, the BAT instrument has detected more than 1000 gamma ray bursts, with an average of 90 GRBs/yr (about 90 \% of them with T$_{90}$ $>$ 2 s).
This large trigger efficiency is due to the exceptional grasp of the instrument  (Fig. \ref{fig_fom}), combined with a strategy to avoid the galactic plane.

\subsubsection{The CZTI instrument from Astrosat mission}
ASTROSAT is a multi-wavelength astronomy mission based on an IRS-class satellite placed on a 650-km, near-equatorial orbit by the Indian launch vehicle PSLV on September 28, 2015, with an expected operating life time of more than five years \citep{Agrawal2004, Singh2014}. 
ASTROSAT embarks five payloads for simultaneous multi-band observations, one of which, called CZTI, is a coded-mask imager with Cadmium-Zinc-Telluride detectors \citep{Vadawale2016,Bhalerao2017}. 
This instrument covers hard X-rays from 20 keV to 200 keV, with a small field of view of 4.6\degree x $4.6\degree$ (FWHM).
Thanks to the thickness of its detectors, above 100 keV CZTI acts as an open detector with a large field of view.
The geometrical area of the detection plane is 976 cm$^2$, providing an effective area of 420 cm$^2$ at 100 keV under normal incidence, when the mask opacity is taken into account. The mass of ASTROSAT is 1505 kg with CZTI weighing $\sim$50 kg.

The detection plane is made of 64 patterned CZT detectors of size \mbox{$39 \times 39$ mm}, with a thickness of \mbox{5 mm} offering a very good detection efficiency, close to 100 \% up to 100 keV, and showing an excellent energy resolution ($\sim$11 \% at 60 keV). 
Each detector has 16 electrodes on each side, making it a small imager with 256 pixels of size \mbox{$2.46 \times 2.46$ mm}, read by two 128-channel ASICs.
The small dimensions of the pixels permits medium resolution imaging in hard x-rays. 
The detection plane is modular, being constituted of four identical and independent quadrants, each made of \mbox{$4 \times 4$} detectors. 
An alpha/gamma source of $^{241}$Am illuminates the whole detector, for in orbit calibration.
The nominal operating temperature is ranging from 5 to 15\degree C, thus facilitating the instrument design. 
The CZTI looks at the sky through a two dimensional coded mask, for imaging purposes. 
Sky images are obtained by applying a deconvolution procedure to the shadow pattern of the coded mask recorded by the detector.

CZTI has detected more than 160 GRBs in 3 years ($\approx55$ GRBs/yr), of which only one was inside the field of view (GRB160325A).

\subsubsection{Some comments on current instruments}
It is quite remarkable that the operating hard X-ray imagers are all based on the same technology: a pixelated CdTe or CZT detection plane with ASICs behind a coded mask. 
In fact coded mask imaging is the only technology offering large fields of view in the hard X-ray range while accommodating detection planes with a geometric area of a few thousand cm$^2$ and tens of thousands pixels. 
It proves highly reliable, as demonstrated by the long lifetime of \textit{INTEGRAL}/ISGRI and \textit{Swift}/BAT.
\textit{Swift}/BAT has shown that, with a large effective area and field of view, these detectors can detect $\sim 100$~GRBs/yr, including events located at very high redshifts up to z = 9.
The existence of patterned detectors with many electrodes on a single crystal allows doing small pixels (\mbox{$\sim1$~mm}), which permit arcminute precision imaging with compact detectors, having the mask placed only a few tens of cm above the detectors.

\begin{table} 		
\caption{Current missions}
\label{tab_current}       
\begin{tabular}{l l l l}
\hline\noalign{\smallskip}
 Mission  & \textit{INTEGRAL}$^a$ & \textit{Swift}$^c$ & \textit{ASTROSAT}$^e$ \\
 Instrument  & ISGRI$^b$ & BAT$^d$ & CZTI$^f$ \\
 \noalign{\smallskip}\hline\noalign{\smallskip}
 Detection  area  (cm$^2$) & 2600 & 5000 & 976 \\
 Field of view (sr) & 0.11 & 1.4 & 0.01{$^1$} \\
 Energy  range (keV)  & [15 - 1000] & [15 - 150] & [20 - 200] \\
 Angular resolution (arcminute)  & 12 & 17 & 18{$^1$} \\
 GRB/yr  & 9.3 & 90 & 55 ($\sim1^1$) \\
\noalign{\smallskip}\hline\noalign{\smallskip}
 Detectors type  & CdTe & CZT & CZT \\
 Number of detectors / pixels & 16384 / 16384 & 32768 / 32768 & 64 / 16384 \\
 ASIC number  & 4096 & 256 & 128 \\
 Operating  temperature  ($^\circ$C) & 0\degree & 20\degree & [5\degree - 15\degree] \\
\noalign{\smallskip}\hline\noalign{\smallskip}
 Orbit height$^2$ (km) & 9,000 $\times$ 150,000 & 600 & 650 / 6\degree \\
 Orbit inclination$^2$ (degrees)  & 52\degree & 22\degree & 6\degree \\
Mission duration (yr) & $\geq 16$ & $\geq 15$ & $\geq 5$ \\
\noalign{\smallskip}\hline\noalign{\smallskip}
\end{tabular}

{$^1$} For GRBs occurring inside the FoV of CZTI.

{$^2$} Orbit height and inclination are given at the beginning of mission's life 

\textbf{References:} {$^a$}{\cite{Winkler2003}}, {$^b$}{\cite{Lebrun2003}}, {$^c$}{\cite{Gehrels2004}}, {$^d$}{\cite{Barthelmy2005}}, {$^e$}{\cite{Agrawal2004, Singh2014}}, {$^f$}{\cite{Vadawale2016,Bhalerao2017}} 
\end{table}

\subsection{Future instruments}
\label{sub_futuremissions}
Several space missions embarking \wfis\ are under development for a launch in the coming years. 

eXTP is an international space mission that will study the state of matter under extreme conditions of density, gravity and magnetism, it is proposed for a launch in $\sim2025$ by a collaboration that gathers institutions of the Chinese Academy of Sciences as well as institutions in several European countries and the United States,. 
The satellite will focus on isolated and binary neutron stars, strong magnetic field systems like magnetars, and stellar-mass and super-massive black holes. 
eXTP will enable the simultaneous spectral-timing-polarimetry studies of cosmic sources in the energy range 0.5-50 keV. 
Its payload encompasses four instruments, among them the Wide Field Monitor (WFM) is a set of 6 coded mask units, with a wide field of view of $50\degree \times 50\degree$ (half sensitivity). 
Each camera has a detection plane area of 180 cm$^2$ made of 4 position-sensitive Silicon Drift Detectors. 
More information on this instrument can be found in Table \ref{tab_future} and in \cite{Hernanz2018}.

THESEUS (Transient High-Energy Sky and Early Universe Surveyor) is a space astrophysics mission developed by a large European collaboration for the ESA Cosmic Vision program, for a possible launch around 2030.
It will contribute to the understanding of the early Universe and to multi-messenger and time-domain astrophysics.
The THESEUS payload is a combination of instruments based on different technologies; for example, a monitor with the lobster-eye telescope technology, will be capable of focusing X-rays in the range $0.3-6$ keV. 
In this paper we briefly present the X and gamma-ray imaging spectrometer (XGIS) designed to detect and localize GRBs in the energy range [2-30]~keV.
XGIS is composed of a set of coded-mask cameras using monolithic X/$\gamma$-ray detectors bars of CsI readout by silicon drift detectors.
XGIS is made of three identical units looking at different regions of the sky, each unit has a field of view of $60\degree \times 60\degree$ (half sensitivity) and a large large detection plane with an effective area of 1024~cm$^2$.
More characteristics of this instrument are given in Table \ref{tab_future} and a detailed description can be found in \cite{Campana2018}.

ECLAIRs encompasses a single unit with a 1024 cm$^2$ detection plane made of 6400 CdTe pixels, much like ISGRI or BAT.
The instrument extends the detection of transients to lower energies, with a low energy threshold at 4 keV. 
ECLAIRs will be launched on the SVOM satellite in 2021 \citep{Wei2016}, and the instrument is now under construction in various French institutes.
The next sections give a detailed description of the instrument and draws some lessons learned during the construction of the detection plane.
A summary of the instrument characteristics is also given in Table \ref{tab_future}.

\begin{table} 		
\caption{Future missions}
\label{tab_future} 
\begin{tabular}{l l l l}
\hline\noalign{\smallskip}
 Mission  &\textit{SVOM}$^a$ & \textit{eXTP}$^c$ & \textit{THESEUS}$^e$ \\
 Instrument  &ECLAIRs$^b$ & WFM$^d$ & XGIS-LE$^f$ \\
\noalign{\smallskip}\hline\noalign{\smallskip}
Detection  area  (cm$^2$) & 1024 & 1080 & 512 \\
 Field of view (sr) & 2.0 & 3.7 & 2.51 \\
 Energy  range (keV)  & [4 - 150] & [2 - 50] & [2 - 30] \\
 Angular resolution (arcminute)  & 86 & 4.2 & 25 \\
\noalign{\smallskip}\hline\noalign{\smallskip}
 Detectors type  & CdTe & SDD & SDD$^g$ \\
 Number of detectors / pixels  & 6400 / 6400 & 24 / $\approx 3~10^{6}$  & 24576 / 24576 \\
 ASIC number  & 200 & 24 & 7680 \\
 Operating  temperature  ($^\circ$C) & $-20$ & $-30$/$-3$ & $-20 / 0$ (TBC) \\
\noalign{\smallskip}\hline\noalign{\smallskip}
 Orbit (km)  & 650 & 550 & 600  \\
 Mission duration  (yr) & 3 & 5 & 3 \\
\noalign{\smallskip}\hline\noalign{\smallskip}
\end{tabular}

\textbf{References:} {$^a$}{\cite{Wei2016}}, {$^b$}{\cite{Godet2014}}, {$^c$}{\cite{Zhang2016}}, {$^d$}{\cite{Hernanz2018}}, {$^e$}{\cite{Campana2018}}, {$^f$}{\cite{Amati2018, Stratta2018}}
\end{table}

\subsection{Some comments on future instruments}
As a conclusion of this short survey, we note that many groups propose ambitious \wfis , demonstrating a continued interest for the detection of hard X-ray transients.
Unlike the instruments currently in operation, however, two instruments use SDDs.
These detectors offer some interesting features, like their sub-millimeter 1D spatial resolution, permitted by closely spaced electrodes, which allows getting arcminute angular resolution in one direction, with distances of only few tens of cm between the mask and the detection plane.
Another interesting feature of SDDs is their capacity for efficient charge transport over cm-size distances.
This permits to measure the other position of an interaction with distant electrodes, using the shape or the drift time of the electron cloud. 
Ultimately, this capability allows reducing the number of electronics readout channels, which increases more slowly than the geometric area of the detection plane.
Overall, SDDs permit building lighter and smaller instruments, although with imaging capabilities restricted to a small energy band that does not extend beyond 30 keV, and to regions far from the galactic plane since the 1D imaging increases significantly the risk of confusion.

Another remark concerns the grasp of these instruments and the size of their detection planes.
Figure \ref{fig_fom} shows that all three projects have a grasp in between the grasps of ISGRI and BAT.
In this respect ECLAIRs and WFM are almost comparable, a factor 2 to 3 below XGIS.
The sensitive area of the 3 instruments is comparable: $\mathrm{\approx1000~cm}^2$ for ECLAIRs and WFM versus $\mathrm{\approx3000~cm}^2$ for XGIS.
The increased grasp is mostly due to the increase of the field of view permitted by the construction of several units looking in different directions.
XGIS, which has the largest grasp, remains however a factor three below \textit{Swift}/BAT.
This may be surprising, according to the need for improved sensitivity pointed in section \ref{sec_detection}. 
There are however several reasons for this choice: first, considering that the Number-Intensity distribution of GRBs flattens at low intensity, it is easier to increase the number of detected GRBs with a larger field of view than with a larger detecting area ; second, \wfis\ are often part of a complex payload, which includes other instruments and they benefit from limited resources.
Third, XGIS is complemented by SXI, a wide field soft X-ray Imager, based on lobster eye optics, which detects long and soft transients and is more sensitive than XGIS for transients longer than ${\approx20~s}$.
Since none of these future instruments will have the grasp of \textit{Swift}/BAT, we consider that it is relevant to discuss here the conditions for the realization of very large detection planes, needed for sensitive \wfis . 
A first step in this direction will be done in Section \ref{sec_discussion}.

Interestingly, various technological choices are shared by the three projects.
This includes a trend to reduce the size and weight of the instruments with smaller pixels and a lower energy range for GRB localization, a high modularity of the detection planes, which are made of small hybridized units regrouping tens to hundreds of pixels read by multichannel ASICs, and a working temperature close to 0\degree C or slightly below, 
It should be pointed that the modularity of the detection plane increases the overall reliability of the instrument because coded-mask imaging is highly resilient to the loss of one or several pixels or even to the loss of a full detection unit. 
Finally, considering the need for many thousands of pixels and the technological evolution of the detectors, the front-end electronics has to become smaller, embedded, with a low power consumption, regrouping many dedicated functions into an unique chip, usually an application specific integrated circuit (ASIC).
Another important point to consider when developing large detection planes is the cost of the detectors; in this respect CdTe remains competitive compared to silicon and can be the best choice for a widely used technology, especially for instruments with an energy range extending beyond \mbox{$\sim 30$ keV}.

Let's now focus on CdTe detection planes and discuss the lessons during the construction of the detection plane for ECLAIRs, called DPIX.


\section{Lessons from ECLAIRs/DPIX}
\label{sec_eclairs}
\subsection{The ECLAIRs instrument from SVOM mission}
The SVOM space mission has been selected by the French and Chinese space agencies to provide the scientific community with precise measurements of gamma-ray burst prompt and afterglow emission, based on a multi-wavelength approach \citep{Wei2016}. 
It carries the ECLAIRs hard X-ray imager, whose X-ray sensitivity will favor the detection of highly redshifted events (z$\ge 5$) and soft GRBs. 
For this reason the \textit{SVOM} team has required ECLAIRs to be sensitive down to 4 keV. 
The telescope, with a good sensitivity in the energy range [4 - 70 keV], is designed to detect transient sources in a large field of view ($90\degree \times 90\degree$ at zero response) and to locate them with an accuracy better than 13 arc-minutes. 
ECLAIRs will deliver almost instantaneously a trigger signal to the platform, but also to ground-based instruments, for rapid follow-up observations. 
The mission launch is planned for December 2021.

ECLAIRs combines a pixelated detection plane with a coded mask which encodes the signal coming from sources in the field of view, allowing to locate them. 
An on-board computer, associated with the detection plane and the mask, will process the recorded detector image (sometimes called a ``shadowgram'') in real time for short periods of integration, and continuously build sky images in which new transient sources will be searched for thanks to a fast deconvolution algorithm.

The detection plane is made up of 200 electrically independent elementary units, each of them encompassing 32 CdTe detectors (4x4 mm$^2$ x1 mm thick) called ``XRDPIXs'', which are mechanically assembled on a cold plate. 
To obtain the lowest electrical noise and to avoid the phenomenon of detector polarization, which reduces the size of the depleted zone, the detectors are cooled to $-20$\degree C, requiring the implementation of a sophisticated thermal control system.
The 200 flight XRDPIX units will be selected among 240 units, according to their performance, homogeneity and stability.


\subsection{The XRDPIX unit}
In this section, we focus on the detection plane by describing the technological choices made to respond to the challenge of the low energy threshold. 
In order to achieve the scientific objectives of the SVOM mission, and to obtain a low energy threshold of 4 keV, uniformly over the whole ECLAIRs camera, various technical issues have been addressed. 
Indeed, the typical performance obtained so far with this type of detectors, is a low threshold of about 15 keV, achieved by \textit{INTEGRAL}/ISGRI and \textit{Swift}/BAT. 

We also describe in this section the hybrid unit XRDPIX, the basic element of the ECLAIRs detection plane, and we outline the lessons learned by designing it. 
The interested reader can also refer to \cite{Ehanno2007, Remoue2008, Remoue2009, Lacombe2013, Lacombe2014, Lacombe2016, Nasser2014}, and \cite{Amoros2018} for extensive discussions of the development process.

The XRDPIX unit is a hybrid, which combines innovative technologies specifically chosen for the project: a CdTe Schottky semiconductor, an already proven detector with excellent characteristics for the detection of gamma-ray photons and more stable at lower temperature. 
This crystal is dense and its compounds have a high atomic number (\mbox{Z = 48} for Cd and \mbox{Z = 52} for Te) and its high resistivity, allowing working at room temperature, even if it was decided to place the ECLAIRs detection plane at low temperature to limit the noise of the detectors due to the thermal agitation and optimize its performance (an average leakage current of \mbox{$\sim 30$ pA} at $-20$\degree C permits a lower energy threshold). 
The CdTe has few negative aspects, such as the widening of peaks above few tens of keV due to the incomplete collection of charge carriers and to the ballistic losses of holes.
Nevertheless it appeared as a very good candidate for the chosen energy range [4 - 150 keV], offering good spectroscopic performance. 
Then, other technologies include: two types of ceramics (thin film and HTCC) acting as substrates respectively for the detectors (which are polarized with a Kovar grid) and for the ASIC chip, which is based on silicon.
The ASIC chip is the IDeF-X ASIC developed by CEA \citep{Gevin2009}, which has been chosen for its high performance in terms of very low ENC: a mean of 60 e- rms at the 4.4 µs peaking time, ensuring an expected energy threshold close to 2 keV. 
Finally, the majority of these components is assembled with a specific conductive glue \citep{Lacombe2018}. 
 
Why a hybrid? this multi-technology component was developed with the purpose of achieving unique performance in sensitivity at low energies. 
Thanks to the experience of ISGRI and BAT, it was decided to use this technology, which is in line with our specific energy range, and shows very low leakage current at low temperature more easily manageable than for other detectors.
The crucial steps in the design of the XRDPIX module were the use of materials with demonstrated low capacitance, and the detailed characterization of the noise generated by the front-end electronics and by the CdTe detectors. 

Then, constraints related to the space mission led us to integrate low consumption and low noise electronics, innovative detectors, high voltage circuits (HV) and weakly dissipative substrates, whose physico-mechanical and thermal characteristics remained compatible with each other. 

\begin{figure}[h!]			
\centering
\includegraphics[height = 6.2cm]{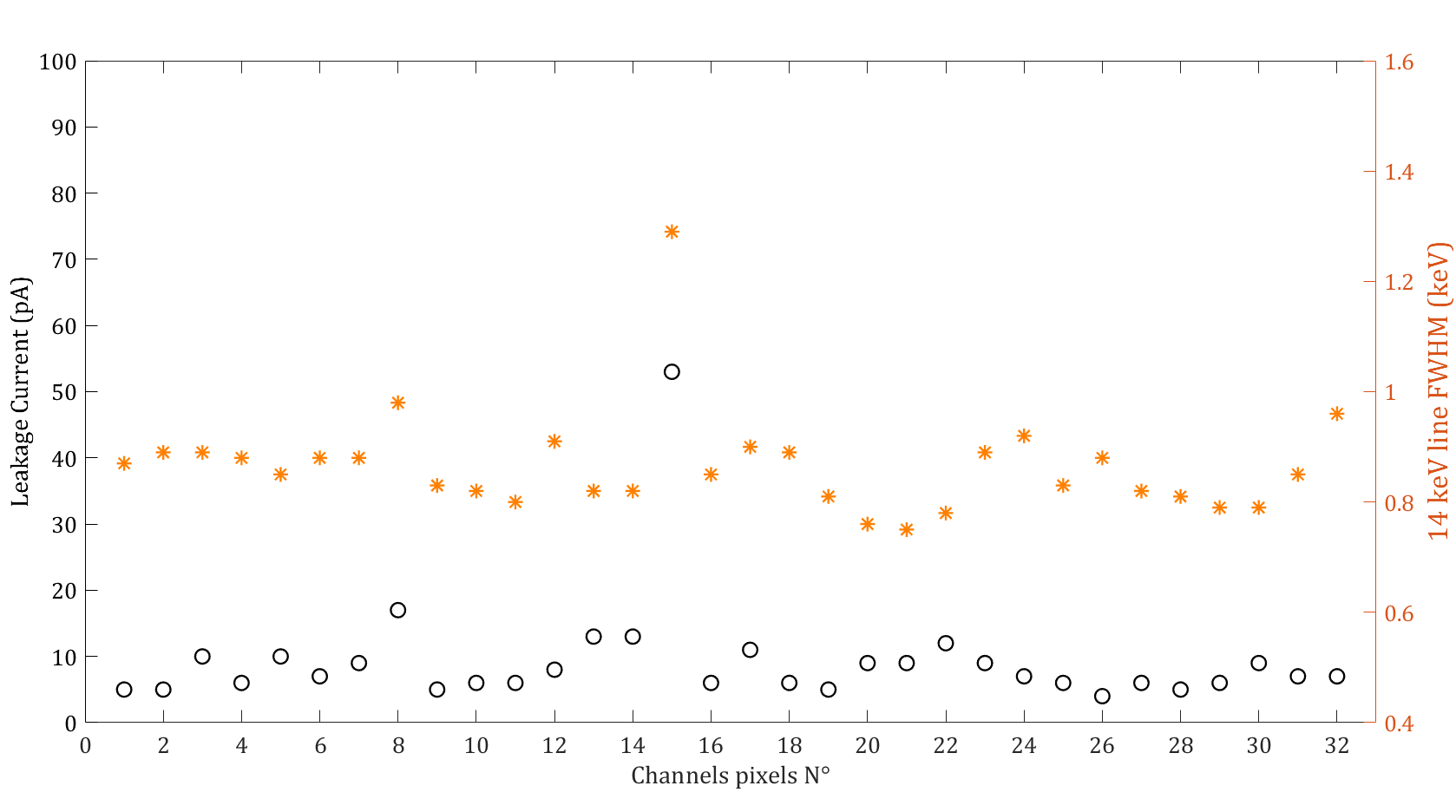}
\caption{Leakage current of 32 pixels biased at $-300$ V (Left Y axis) - 14 keV line FWHM for the configuration ($-300$ V; $2.6 \mu s$) - XRDPIX unit XR133.}
\label{fig_coupling}
\end{figure}

Finally, a modular solution was adopted for the development and characterization of the hybrid XRDPIX, by choosing to develop it in two sub-elements, so called ``Detectors Ceramics'' and ``ASIC Ceramics''. 
This modularity allows to get a great flexibility on the units characterization, to select a low noise and low consumption ASIC with a 32-channel multiplexing, to obtain reduced overall dimensions, to reduce waste, and finally to optimize the performance  by coupling  ASIC channels to detectors individually. In the end this resulted in a reduction of the electronic noise, permitting to detect photons below the nominal threshold of 4 keV. 
As shown in figure \ref{fig_coupling}, the homogeneity of the spectral resolution and the low energy threshold are obtained by avoiding the association of a high leakage current detector with an ASIC channel showing a too high ENC (typically higher than 70 e- rms). 
This key step in the development of the detection plane further justifies the choice of modularity.

\begin{figure}[h]				
\centering
	\includegraphics[height = 5.2cm]{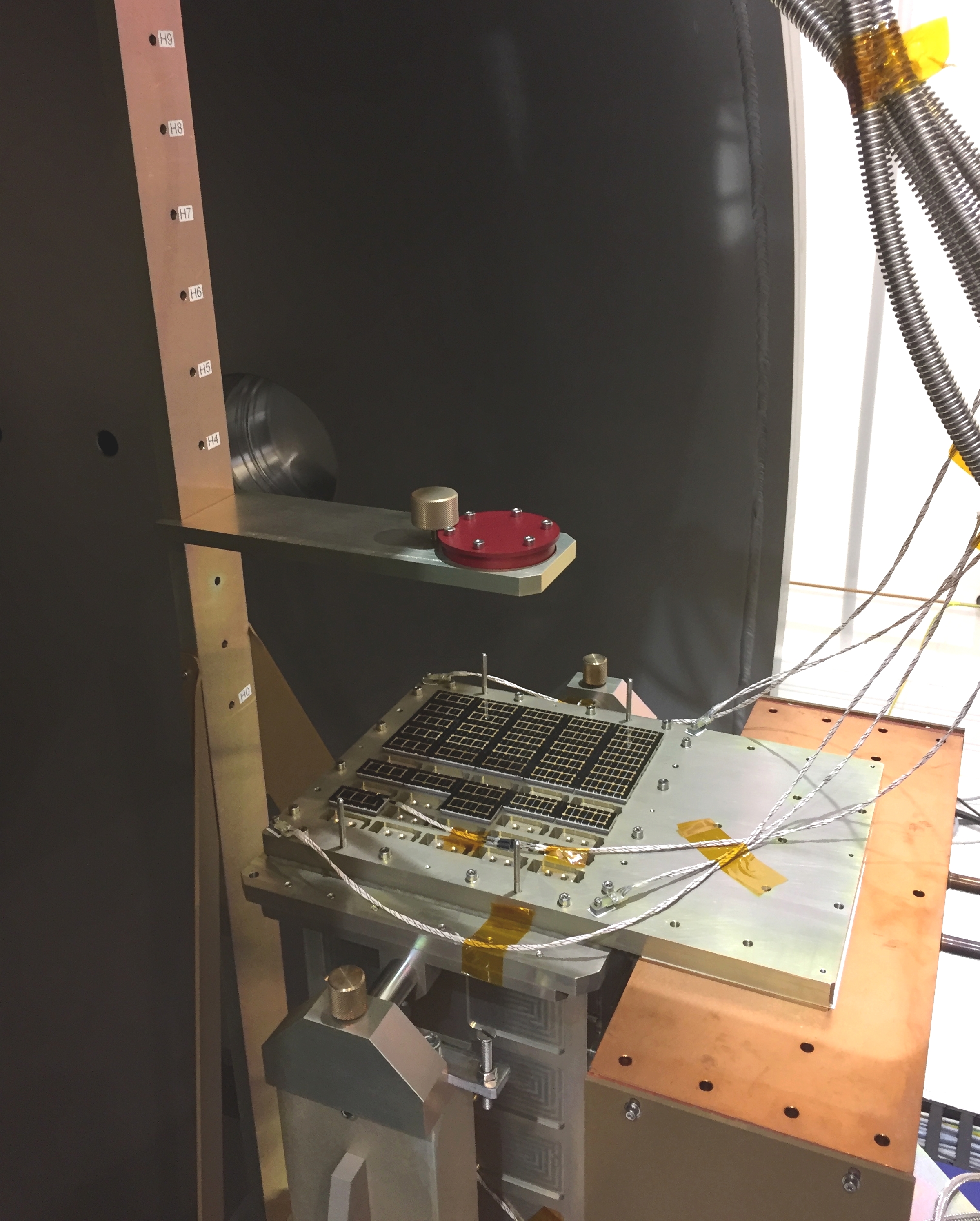}\hfill
	\includegraphics[height = 5.2cm]{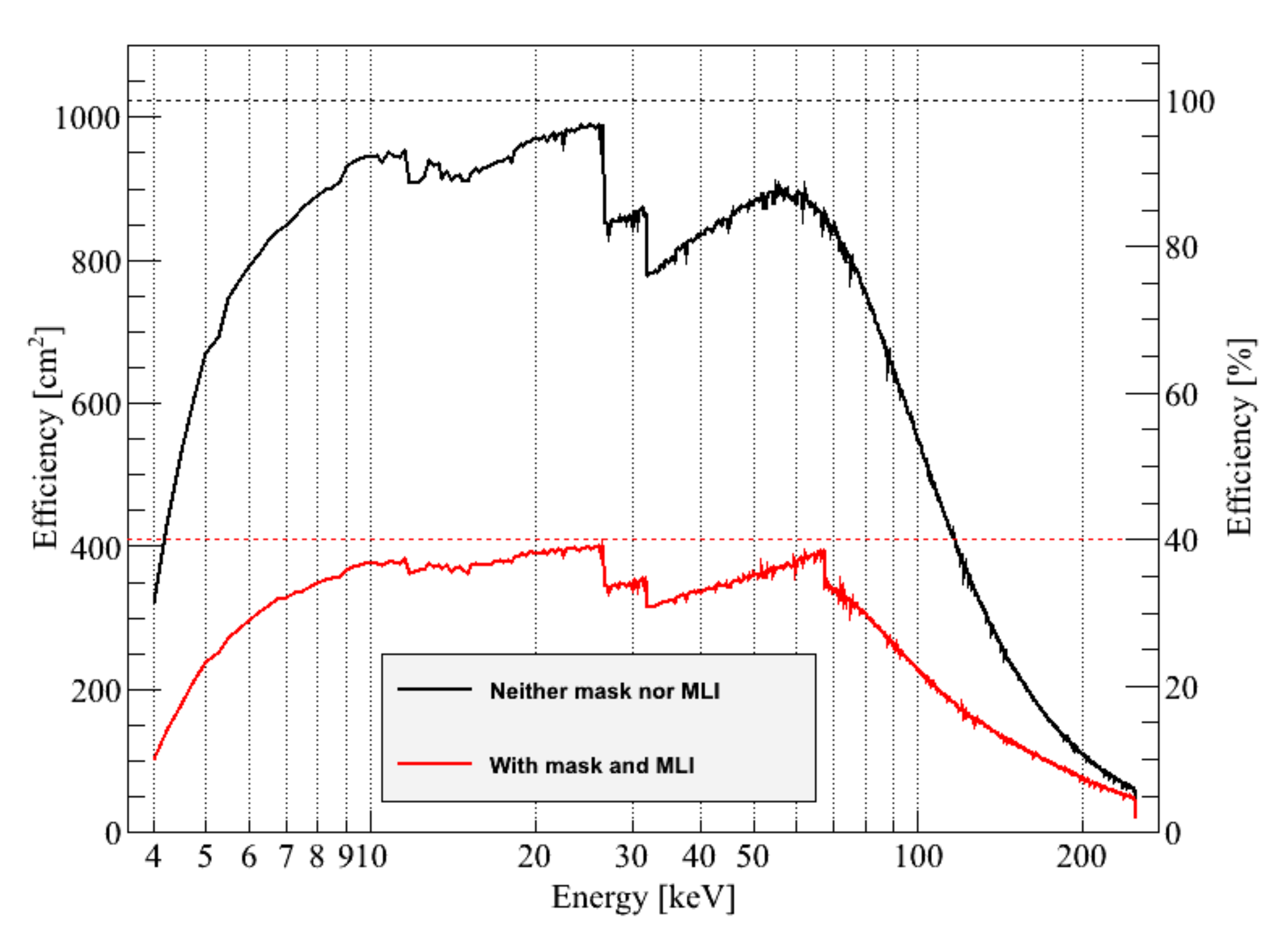}
\caption{Left: Prototype camera of the ECLAIRs detection plane under test in the vacuum chamber. Right: Effective area at peak energy for the detection plane and of the instrument, from Geant 4 simulations (P. Sizun, \textit{SVOM} internal report).}
\label{fig_proto} 
\end{figure}

\begin{figure}[h]				
\centering
\includegraphics[height = 8.8cm]{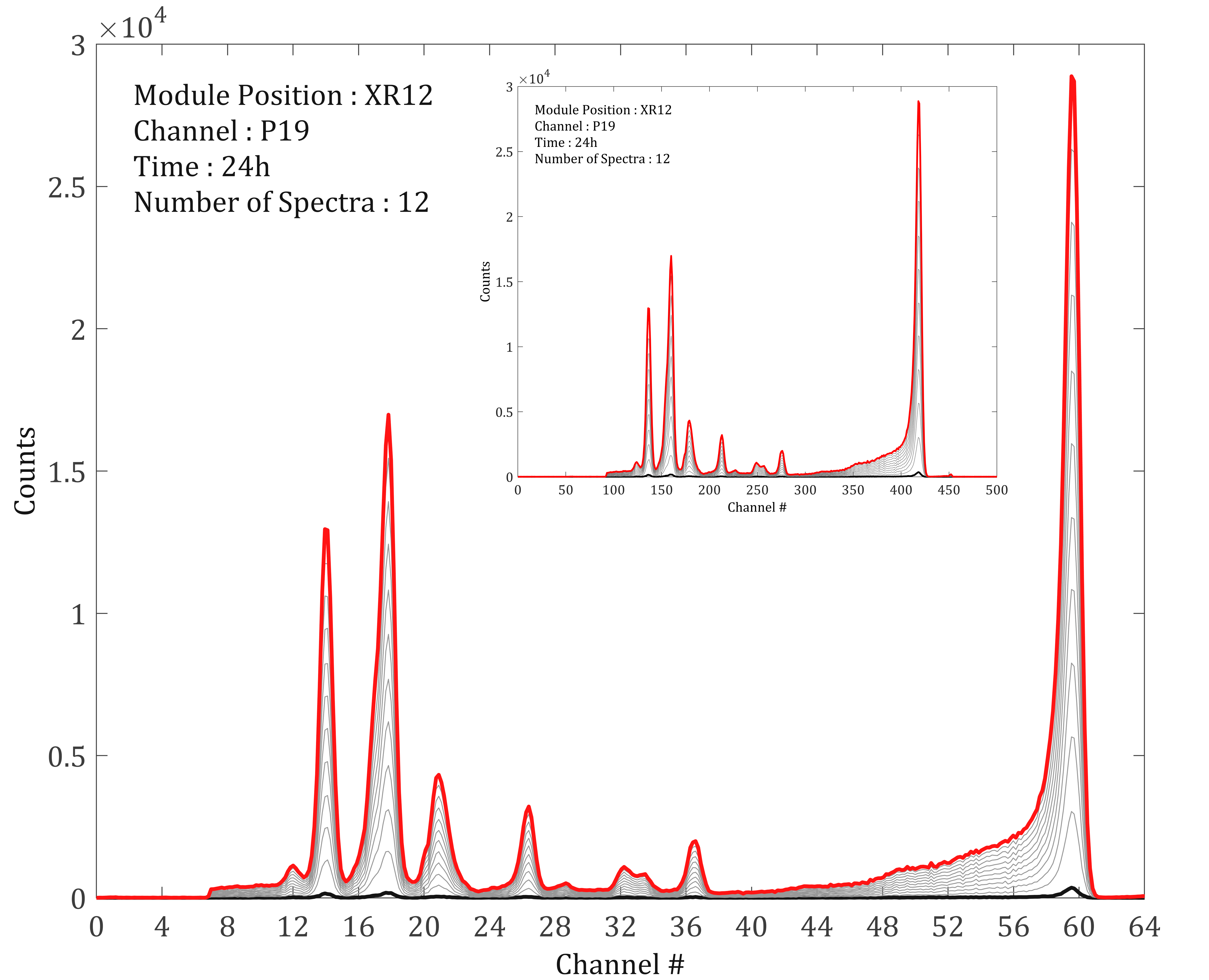}
\caption{$^{241}$Am spectra after calibration acquired during 24h at $-20^{\circ}$C, with a $-300$ V bias voltage (inset graph: spectrum before energy calibration) - XRDPIX unit XR029 - Channel pixel P12 - Position on prototype X12 - One spectrum every two hours.}
\label{fig_QE} 
\end{figure}

\subsection{The ECLAIRs optimization and expansion}
In the event of the development of future instruments (for example an instrument as ECLAIRs to be developed in a shorter mission time), several lessons can be drawn from this experience. 
In general, it would be recommended to opt for a new strategy of manufacturing, design and testing. 
First of all, it is not necessary to test all the detectors individually before gluing them on the detector ceramics (we characterized around 14000 CdTe detectors for a selection of 6400) but it is sufficient to test the 32 detectors matrices ('Detectors Ceramics'). 
This is justified by the measure of an increase of the leakage current after having glued the detectors on their substrate. 
Then we propose adopting a different quality policy (for example a batch policy), by choosing to test only a fraction of the 'Detectors Ceramics' (e.g. 30 \% ). This is justified by the excellent homogeneity of the crystals provided by the ACRORAD company. Moreover, the manipulation of unitary CdTe detectors increases the risk of their degradation. Then, given their high stability in performance (particularly in terms of leakage current), it seems that a batch strategy is more suited to the timing of projects involving tens of thousands detectors.

A simplification of the ASIC chip is also desirable, removing some secondary or even optional features, which can cause additional noise problems and thus impact the low energy threshold performance. 
For example, a single value of peaking time may be adjusted, without the need to program it, inducing digital signals which may cause cross-talk. 
It is also important to optimize the ASIC connections to get input lines with more uniform length (e.g. using two sides of the chip), and to enlarge the electrical shielding around the 'ASIC Ceramics'.

\begin{figure}[h]				
\centering
\includegraphics[height = 8.8cm]{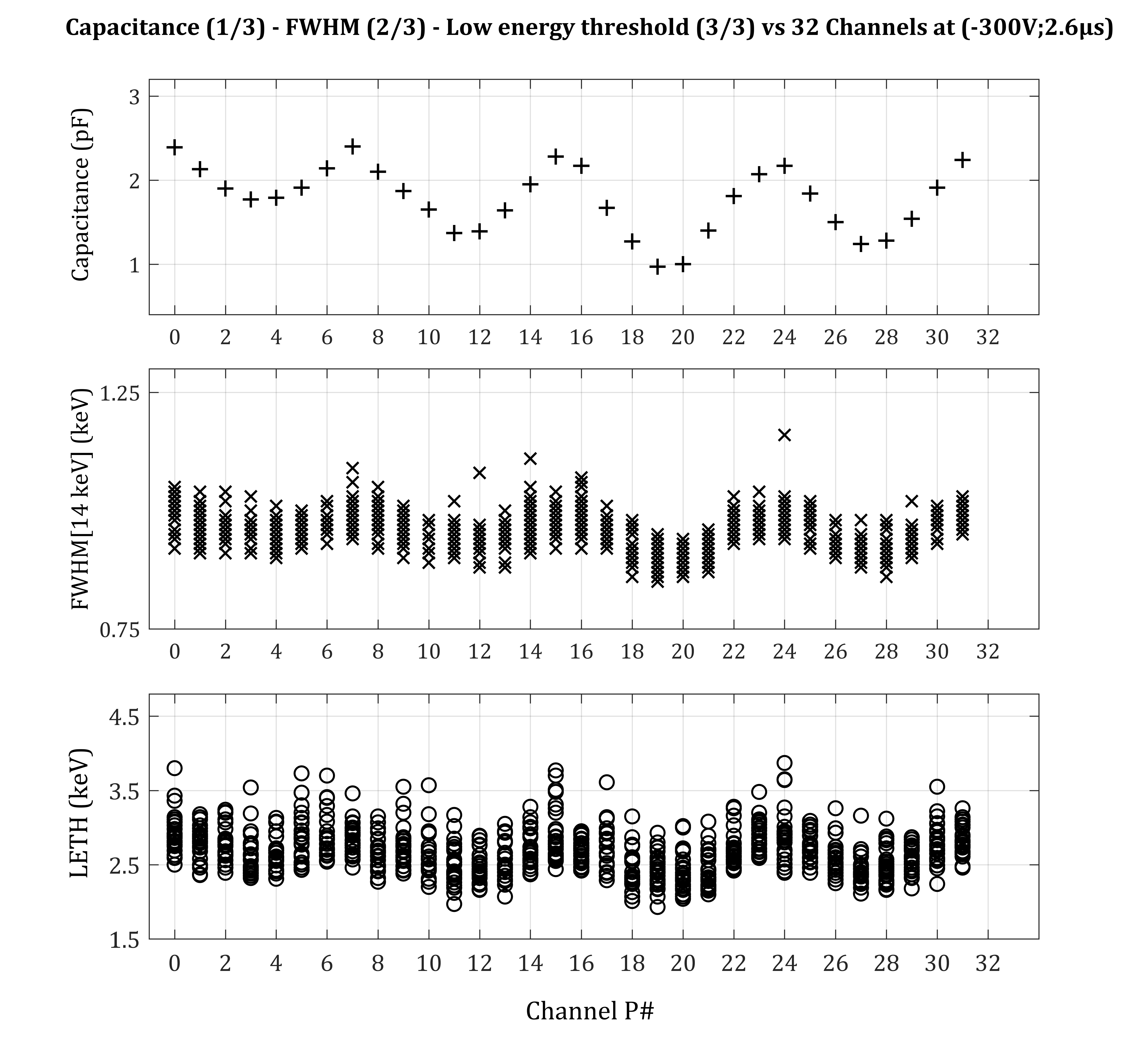}
\caption{Impact of the capacitance of the readout circuit on the final performance of the detectors - Top: Capacitance of the 32 channels of one ``Detectors Ceramics'' - Middle : FWHM at 13.9 keV for the ($-300$ V; $\mathrm{2.6 \mu s}$) configuration - Bottom: Low energy threshold for the 32 channels of 45 XRDPIX modules in the ($-300$ V; $\mathrm{2.6 \mu s}$) configuration.}
\label{fig_capa} 
\end{figure}

To control the electrical noise, it is necessary to focus either on a completely analog ASIC chip, by removing in this case the clock signal ('strobe'), or a numerical ASIC chip that processes directly the output signal of detectors, by reducing the length of tracks.
Finally, it is necessary to study the homogeneity of the routing of detectors output wires with the routes of digital signals and power supply. 
As a matter of fact, the equivalent capacitance values of a circuit, whatever their material (PCB or ceramics) could shape the performance of detectors channels, as can be seen in figure \ref{fig_capa}. 
Moreover, we note that PCB (printed circuit) is less favorable compared with ceramics. 
Indeed, ceramics provides better thermal compatibility and the grip with pixels is more controlled. These improvements will permit to obtain an energy threshold of $\sim $ 10 keV with an instrument working at room temperature, thus avoiding the need for cooling (but not the need for temperature stabilization). 

Finally, to build an instrument more performing than ECLAIRs with the same architecture, it is needed to improve the robustness of XRDPIX unit, with a protection around the fragile CdTe crystals, such as a resin, considering manufacturer feedback.

In summary, modularity is essential. 
The simplification of the test and validation phases is desired. 
Avoiding mixing analog signals with digital signals is required. 
Nevertheless, this technology remains complex to implement for making several thousands of square centimeters. We address this issue in the next section, showing that it can be considerably alleviated with the use of patterned CdTe detectors.

\section{Large detection planes for future projects}
\label{sec_discussion}

\subsection{General requirements}
\label{sub_sensireq}
In this section we tackle the design of large detection planes for future coded-mask imagers.
We justify this discussion by the scientific need for highly sensitive detectors for short X-ray transients.
Moreover, considering the long history of successful instruments working with CdTe, the continued progress in the realization of CdTe detectors, and our own expertise, we restrict the discussion in this section to detection planes based on CdTe detectors (except for a short digression in section \ref{sub_detector}.

\subsubsection{Detection area}
\label{sub_area}
As explained in section \ref{sec_detection}, the need for hard X-ray imagers with large detection areas will remain strong in the coming years.
As shown in Tables \ref{tab_current} and \ref{tab_future}, space instruments with pixelated sensitive areas equal or larger than \mbox{0.1 m$^2$} (1000 cm$^2$) are currently built. 
The record being held by \textit{Swift}/BAT, with an area of \mbox{5200 cm$^2$} encompassing 32768 detectors of CdTe read out by 256 ASICs.
While it seems difficult to overcome this size for detection planes with individual pixels of CdTe (for budgetary, complexity, mass and volume reasons), the progress of detector technology opens new avenues for the construction of large area detection planes with reasonable resources.
Before addressing some of these opportunities in section \ref{sub_detector}, we briefly comment some important features that need to be considered in the design of large detection planes.

\subsubsection{Energy range}
\label{sub_energy}
Unlike polarimetry or spectroscopy, imaging does not require a broad energy range since it is much more important to get many photons than to get photons with vastly different energies.
A decade in energy is usually acceptable since it provides a crude spectral indication (e.g. a PL index), which is important for source classification (along with the temporal evolution), while avoiding the need for an electronics with a large dynamic range.
Considering the need for photons of coded-mask imaging (multiplexing the sky pixels on the detection plane requires a minimum number of photons to detect a point source without ambiguity), low energies are preferred, typically of the order of tens of keV.
Working in this energy range permits limiting the mass of the instrument, which may have mask elements and shield that are moderately thick (e.g. 0.3 mm of tantalum absorb $\geq 98 \%$ of X-rays below 80 keV).
On the low energy side, going to keV energies raises two additional challenges. 
First, the transparent mask elements must be truly transparent, preventing the use of any support structure and calling for self-supported coded-masks, which are very difficult to build, especially for large instruments.
Second, the detectors require a lower operating temperature, complicating the overall thermal system.
We thus recommend to go down to a few keV only if this is strongly required by the scientific objectives of the mission, otherwise an energy threshold of the order of 10~keV is more easily accessible.

\subsection{Angular resolution and number of pixels}
\label{sub_pixels}
The angular resolution of coded mask imagers is roughly proportional to the ratio x$_{\mathrm d}$/H, where x$_{\mathrm d}$ is the size of mask elements and H is the height of the mask above the detectors.
It is usually demanded to wide-field imagers to have sub-degree angular resolution. 
This accuracy, while it is insufficient to identify the host galaxies of extragalactic transients, is sufficient to permit follow-up observations with narrow-field X-ray or optical telescopes that can localize the afterglow of the source with arcsecond precision.

\textit{INTEGRAL}/ISGRI, \textit{Swift}/BAT, \textit{SVOM}/ECLAIRs all have pixels of the same size \mbox{$4 \times 4$ mm$^2$}, but with different choices in terms of field of view and angular resolution. 
\textit{INTEGRAL}/ISGRI offers the best angular resolution (12 arcminutes, with x$_{\mathrm d}$ = 11.2~mm, and H = 3.2~m) and the best imaging efficiency, albeit with a relatively small field of view.
\textit{Swift}/BAT offers a tradeoff between angular resolution, field of view (17 arcminutes, with x$_{\mathrm d}$ = 5~mm, and H = 1.0~m), and imaging efficiency (\mbox{72\%}).
Being limited by the overall space available for the instrument, \textit{SVOM}/ECLAIRs offers limited angular resolution (85 arcminutes, with x$_{\mathrm d}$ = 11.4~mm, and H = 0.46~m) across a large field of view, with the same very good imaging efficiency (85\%) as ISGRI.
Both \textit{INTEGRAL}/IBIS and \textit{Swift}/BAT are large instruments, weighting respectively 677~kg and 200~kg, \textit{SVOM}/ECLAIRs is significantly smaller and lighter, weighting only $\approx85$ kg.

Considering the volume constraints for space instruments, it is suitable to reduce the size of mask and detector pixels, in order to keep arcminute localization accuracy with a coded-mask located only a few tens of cm above the detectors.
Using smaller detector pixels allows a better sampling of the shadowgram, i.e. a better imaging efficiency, and permits to use smaller mask pixels for a better angular resolution.
While it is possible in theory to go to sub-millimeter size pixels, this would require more than $10^5$ detectors to pave a detection plane of only 1000 cm$^2$.
Even if an ASIC is used to read several tens of detectors, this is an important constraint.
The choice of \textit{INTEGRAL}/ISGRI, \textit{Swift}/BAT, \textit{SVOM}/ECLAIRs was instead to design instruments with rather large individual detectors (\mbox{$4 \times 4$ mm$^2$}), to favor sensitivity over localization accuracy.
The design of the readout electronics for such a large number of pixels appears as a crucial issue for the construction of large detection planes, this issue is discussed in more details in section \ref{sub_detector}. 

\subsection{Modularity}
\label{sub_modularity}
A significant feature of the detection planes of BAT, ISGRI and ECLAIRs is their modularity.
With so many detectors it is essential to adopt a modular approach, dividing the detection plane into hundreds of identical units.
In view of the interest for very large detection planes in the future, it seems wise to extend this approach to the construction of the instrument itself, by making various detection modules (units) working together, each with a detection area of several hundred to one thousand cm$^2$.
For each of them, we note that after a generation of large instruments, like ISGRI or BAT, future projects propose instruments made of several modules with a detection area of about \mbox{1000~cm$^2$}.
This strategy, which has been adopted by \textit{eXTP}/WFM (with 6 modules) and \textit{THESEUS}/XGIS (with 3 modules), enforces the overall reliability of the instrument, and avoids making ``giant'' instruments, which may face problems of alignment, stability and have complex AIT/AIV.
It also makes it easier to control the homogeneity of the detection planes and the instrument performance.
Finally, this approach offers additional tradeoffs between sky coverage (with modules pointing at different directions) and sensitivity (with modules pointing at the same direction), but it requires detailed simulations in the design phase of the instrument to optimize the geometry as a function of the scientific objectives.

\subsection{Choice of detectors}
\label{sub_detector}
We discuss here three types of detectors which have been used or which are planned for future hard X-ray imagers: CdTe, silicon drift detectors (SDD) and scintillators read by SiPM.

\subsubsection{CdTe and CZT}
\label{ssub_cdte}

CdTe and CZT detectors have a long history of successful operation in space for X-ray astronomy. 
The detection planes of \textit{INTEGRAL}/ISGRI, \textit{Swift}/BAT, \textit{SVOM}/ECLAIRs use the same technology, based on individual pixels of CdTe (or CZT) of size \mbox{$4 \times 4$ mm$^2$}.
After 16 years of operation more than 90\% of the 16k detectors of ISGRI remain operational, and 65\% of the 32k detectors of BAT were operational ten years after its launch in 2004 \citep{Lien2014}.
\textit{ASTROSAT}/ CZTI, on the other hand, uses larger and thicker detectors (\mbox{$39 \times 39 \times 5$ mm$^3$}) with a continuous anode on one side and a patterned cathode divided into \mbox{$16 \times 16 = 256$} pixels on the other side. 
The cathode is readout by two 128-channel ASICs.
The technology of CdTe detectors has been evolving quickly in recent years, with the implantation of Schottky electrodes, which reduces the dark current significantly, and with the manufacturing of large detectors with patterned electrodes, which provides sub-millimeter spatial resolution over surfaces of tens of square centimeters. 
These technological developments are very interesting in the context of the construction of large detection planes, especially when we consider the development of double sided CdTe strip detectors (CdTe-DSD, e.g. \cite{Ishikawa2010}), which permits to get a number of output channels significantly smaller than the number of pixels (typically 2$\times$N channels for N$^2$ pixels).
These developments are accompanied with progress in the integrated readout electronics, allowing to reach energy thresholds below 10 keV at room temperature.
Finally the sub-millimeter pitch of CdTe-DSD detectors permits to construct compact detection units, with the mask located $~10$~cm above the detectors.
It is thus clear that large CdTe-DSD detectors are excellent candidates for large detection planes, at the condition to have a moderate temperature regulation (typically below 1$^\circ$C), to avoid large gain shifts. Another point that should be mentioned is the exposure of CdTe detectors to increasing fluences of ionizing radiation, because they can seriously affect their spectroscopic performance and operation.

\subsubsection{Silicon drift detectors}
\label{ssub_sdd}

This technology offers sub-millimeter pixels and a reduced energy range due to the low atomic number of silicon. 
SDDs have been used in space (e.g. Nicer), but not for large detection planes, even if they are the baseline for very large detection planes of various future missions (e.g. LOFT, eXTP).
Both \textit{eXTP}/WFM and \textit{THESEUS}/XGIS propose wide-field hard X-ray imagers based on SDD detectors.
The thermal requirements of SDDs are comparable with those of the CdTe: operation at room temperature or slightly lower and moderate thermal control.
Like CdTe, they are good candidates for the realization of future large detection planes, however with an energy range limited to energies below 50 keV and with less experience in space. However, they are less fragile, especially during handling phases (during testing for example) and their assembly, and working at lower voltages remains an advantage.

\subsubsection{Scintillators}
\label{ssub_scint}
The recent years have seen a renewed interest for scintillators, whose light emission can now be read by photodiodes, like the silicon photomultiplier (SiPM \citep{Buzhan2003}), instead of photomultipliers, which are bulky and fragil.

SiPM are made up of a large number of independent silicon micro-pixels (typically between 20 and $50\mu m$), operating in Geiger mode. 
The photon detection efficiency (quantum efficiency) of SiPM detectors is slightly better than conventional PMTs, around 50\% at maximum, and the gain is equivalent, of the order of 10$^6$. 
Moreover, SiPM present smaller gain fluctuations, as a function of temperature or polarization voltage, than other similar detectors such as an avalanche photo-diode. 
However, where PMTs exhibit a very low dark count rate, that of SiPM is very large at ambiant temperature so that cooling is often required. On the other hand SiPM have a significant advantage over PMTs with their moderate bias voltage (30-60 V) especially for space applications where voltages near 1 kV are always a concern. 

The direct conversion of gamma photons into electron-hole pairs in semiconductors allows a better energy resolution and a better intrinsic spatial resolution. 
The number of charges created in semiconductors is greater than the number of charges created by the photo-cathode of a PMT, coupled to a scintillator. 
For example, a 122 keV photon that interacts in the NaI detector produces about 1,500 photo-electrons, versus 27,500 pairs of electron-hole charges with the CdTe, which needs about 4.43 eV to create a pair. 
This smaller statistical noise explains the excellent energy resolution of semiconductors. 
Besides, the electric field applied between the electrodes of the semiconductor let the charge carriers displacement be directed into one direction, thus improving the quality of the location of photons, unlike the scintillator, where the light emission is isotropic and scatters until it reaches the SiPM. 

New inorganic scintillating materials are studied with the aim of improving the energy resolution and the spatial resolution intrinsic to the detection system. 
A scintillator crystal, ideal for space gamma-ray astronomy, must have high Z components, be dense, fast, providing a good light response and emitting optical photons at wavelengths adapted to the photo-cathode of the SiPM. 
Many crystals are studied for space instruments, the most promising is probably the LaBr3 scintillator; indeed, 5 mm of LaBr3 absorbs 80 \% of photons at 140 keV. 
It is also very fast with a decay time of $26$~ns, and its light response is 1.65 times larger than NaI, making it possible to have an energy resolution of 6.6 \% at 140 keV. 
However, serious disadvantages remain in its use, such as a permanent background activity due to lanthanum (La) which is a natural radioisotope, and an hygroscopic material requiring waterproof encapsulation.

While scintillators are good candidates for the realization of large detection planes, a drawback is the spatial resolution on the detection plane, which is typically a few mm because the light output is read out by centimetre size PMTs  \citep[e.g.][]{Paul1991} and because the energy deposited in the crystal is mediated by light photons, which can propagate over long distances in the crystal .
With such a spatial resolution, the coded mask must be typically located 1 meter above the detection plane in order to get localizations with a precision of few arcminutes, preventing the realization of compact instruments.
However, this situation may evolve with the availability of SiPM, which allow a better spatial resolution at the expense of a larger number of readout channels. This comparison between different detector technologies should not overshadow the radiation problems. Indeed, SiPMs, for example, do not seem to endure some types of particles as protons, which will significantly affect the level of their dark current.

\subsubsection{Summary}
\label{ssub_summary}
In summary, it appears that CdTe remains a serious candidate for the realization of large detection planes in the future. 
This observation is based on the long experience gained with these detectors, their radiation hardness, and the progress made in manufacturing Schottky detectors with patterned electrodes and low-noise ASICs.

\section{Conclusion}
\label{sec_conclusion}

The development in the coming years of several instruments devoted to time domain astronomy (e.g. LSST or SKA) and multi-messenger astronomy (LIGO and VIRGO, ICECUBE, KM3NeT), calls for the continuous survey of the hard X-ray sky with improved sensitivity in the future.
The need to explore the poorly known subluminous GRB populations, like GRB~170817 located at 47~Mpc or GRB~980425 located at 37~Mpc (Fig. \ref{fig_grbvz}), will require large detection planes, typically $\ge 1$~m$^{2}$, as explained in Section \ref{sec_detection}.
Until now, only \textit{Swift}/BAT had such a large detection area, with 32768 CZT detectors, which has permitted significant discoveries, like the detection of GRBs with redshifts up to 9 and the detection of few subluminous GRBs, like GRB~060218 located at 147~Mpc (e.g. \citet[][]{Campana2006, Pian2006, Soderberg2006, Liang2007}).

Based on a short discussion of current and future coded mask hard X-ray imagers (section \ref{sec_status}), and on our experience with ECLAIRs (section \ref{sec_eclairs}), we summarize below the conditions required for the construction of large CdTe detection planes. 
The main qualities of CdTe detectors are their radiation hardness, their high atomic numbers, their high density, the possibility to manufacture them in large quantities, and their low noise at room temperature.
These qualities explain why CdTe (or CZT) detectors have an excellent record of their use in space, with a total detecting area of currently operational detectors reaching nearly 1~m$^2$ (Table \ref{tab_current}).

First, future instrument builders should carefully review the progress in the field of CdTe detectors before choosing the type of detectors they will use. 
For instance, Schottky type detectors, which offer a very low leakage current, permit operating the detectors at room temperature with excellent performance in terms of resolution and energy threshold. 
Furthermore, patterned electrodes permit making detectors with sub-mm pixels, allowing the construction of compact instruments that offer a good angular resolution of a few arcminutes with a coded mask located only a few tens of cm above the detectors. 
The need for large detecting areas with sub-mm spatial resolution results in detection planes with many $10^{5}$ pixels, a number which is problematic if each pixel is read individually. 
The production of double sided CdTe strip detectors \citep[e.g.][]{Ishikawa2010} solves this problem, since for these detectors the number of readout channels grows as the square root of the number of pixels.
The thickness of the detectors must be adapted to the energy range of the instrument. It is recommended to avoid too thick detectors, which require large high voltages and may have a low collection efficiency for the holes.
We conclude that the advances in CdTe detector manufacturing opens the way for significant improvements with respect to current (\textit{INTEGRAL}/ISGRI, \textit{Swift}/BAT, \textit{ASTROSAT}/CZTI) or planned (\textit{SVOM}/ECLAIRs) instruments.

The second point concerns the modularity of instruments involving $\approx 1$~m$^{2}$ of detectors. 
As explained in Section \ref{sub_modularity}, the modularity of the detection plane naturally leads to the instrument modularity, with several modules having $\approx 1000$~cm$^{2}$ of detectors. 
This strategy eases the realization of the instrument, since the modules can be built and tested in parallel, it increases the overall reliability, and it avoids constructing a giant instrument, with all the problems of handling, assembly integration and testing accommodation on the satellite and uniformity of the detection plane.
It also impact the spare model philosophy, reducing greatly the spare model cost.
We note that after a series of single instruments (\textit{INTEGRAL}/ISGRI, \textit{Swift}/BAT, \textit{ASTROSAT}/CZTI), and (\textit{SVOM}/ECLAIRs), the choice of modularity has been adopted for \textit{eXTP}/WFM and \textit{THESEUS}/XGIS.
Modularity is easy to implement when the goal is to increase the field of view, with modules pointed in different directions. 
Its implementation is more complex when the goal is to increase the sensitivity with all modules pointing at the same field of view. 
In this case, the modules have to work in synergy on board in order to get the full potential of the large sensitive area.
The detection of transients needs to be done on a sky image combining the sky images from the different modules, requiring large computing resources on board.
We nevertheless believe that the benefit of modularity is larger than the additional complexity resulting from the need to combine sky images on board.
Finally, when dealing with large sensitive areas, it is essential to simplify the instrumental design, to work at room temperature (or close to), to adopt an industrial approach for the realization of the detection units, and to simplify the testing of the elements. This is made possible thanks to the resilience of coded mask instrument whose performance is not significantly inspected by a few percent of inoperative detectors.

While there are strong scientific needs for sensitive imagers with large detection area for the exploration of the transient high-energy sky, and the technology is available, we believe that significant efforts are required before we can propose instruments with a detection area larger than 1~m$^{2}$, with credible mass, volume and financial budgets. This article is part of this effort, outlining some key points to be addressed in the design of large hard X-ray imagers and trying to convince the reader that CdTe detectors possess significants assets in this respect. 



\bibliographystyle{aa} 
\bibliography{biblio-gammadet} 

\newpage

\end{document}